\documentclass[twocolumn,superscriptaddress,floatfix,showpacs]{revtex4-1}
\pdfoutput=1
\usepackage{graphics,amssymb,amsmath,epsfig,color}
\usepackage{graphicx}

\newcommand{\be}{\begin{equation}} \newcommand{\ee}{\end{equation}}
\newcommand{\bea}{\begin{eqnarray}} \newcommand{\eea}{\end{eqnarray}}
\begin{document}

\title{On the Continuum Time Limit of Reaction-Diffusion Systems}
\author{Peter Grassberger} \affiliation{JSC, FZ J\"ulich, D-52425 J\"ulich, Germany}
         \affiliation{Max Planck Institute for the Physics of Complex Systems, N\"othnitzer Strasse 38, D-01187 Dresden, Germany}
\date{\today}

\begin{abstract}

The parity conserving branching-annihilating random walk (pc-BARW) model is a 
reaction-diffusion system on a lattice where particles can branch into $m$ offsprings 
with even $m$
and hop to neighboring sites. If two or more particles land on the same site, they 
immediately annihilate pairwise. In this way the number of particles is preserved modulo
two. It is well known that the pc-BARW with $m=2$ in 1 spatial dimension has no phase 
transition (it is always subcritical), if the hopping is described by a continuous time 
random walk. In contrast, the $m=2$ 1-d pc-BARW has a phase transition when formulated 
in discrete time, but we show that the continuous time limit is 
non-trivial: When the time step $\delta t\to 0$, the branching and hopping probabilities
at the critical point scale with different powers of $\delta t$. These powers are different
for different microscopic realizations. Although this phenomenon
is not observed in some other reaction-diffusion systems like, e.g. the contact process, 
we argue that it should be generic and not restricted to the 1-d pc-BARW model.

\end{abstract}

\maketitle

\section{Introduction}

It is well known that the short-distance behavior of relativistic quantum field theories 
is in general anomalous. For instance, Abelian gauge theories like QED have diverging bare 
charges while the coupling constants of non-Abelian gauge theories like QCD converge to 
zero at short distances. In simulations of non-linear quantum field theories like QCD this 
is usually taken into account by making space-time discrete and letting the lattice constant 
tend to zero after the calculation.

In non-relativistic reaction-diffusion systems one might a priori expect something similar,
but there it is of course extremely natural to work on spatial lattices, so the problem is 
considered as much less fundamental. Indeed it is often taken for granted that spatial 
discretization is sufficient to render any reaction-diffusion model well defined mathematically.
This is e.g. true for models in the Reggeon field theory \cite{Brower} universality class. 
As shown in \cite{grass78}, Reggeon field theory describes a reaction-diffusion 
system which can be either realized as a process discrete in space and time, 
in which case it is known as directed percolation \cite{cardy80}, or on discrete space 
with continuous time, called `contact process' \cite{marro05}. 
Indeed, the contact process can be seen as the limit of a particle process in discrete time with 
hopping diffusion, branching, annihilation, and spontaneous decay where all probabilities are 
$O(h)$ with $h\ll 1$, and physical time proceeds also slowly so that $t$ is $h$ times the 
number of iteration steps.

In the present note we point out that things are not always so simple. Take a 
diffusion process on a spatial lattice, where the hopping rate is $h \ll 1$. This 
by itself would correspond to a random walk with $R_n^2 \propto hn$. In order to have a 
finite diffusion constant $D$ independent of $h$, we then have to define physical time as 
$t\propto hn$. Add now some reaction(s) which conceivably can lead to qualitatively new 
behavior(s) with one or more  critical points. Is it still true that the critical points 
are obtained when all reaction rates also scale $\propto h$? We will give an example where 
this is not the case.

The model we study here is the one-dimensional branching-annihilating random walk (BARW) with 
two offsprings at each branching event and with pairwise annihilation \cite{GKV,takayasu92}. 
Since both branching 
and annihilation change the number of particles by an even amount, the total number of 
particles is conserved modulo 2. This is also called the parity preserving BARW (pc-BARW).
It was proven rigorously by Sudbury \cite{sudbury90} that this model is always subcritical, if 
treated in continuous time. This is often taken as evidence that the model is always subcritical 
also when treated in discrete time \cite{Jensen94,takayasu92,Takay92}, and more complicated 
models were used to study the corresponding universality class \cite{Jensen94,Menyhard,zhong95}. 
We will see that this is not true.
More concretely, we will study two versions of the discrete time 1-d pc-BARW. In model A 
hopping and branching reactions are applied alternatingly. First, all particles can hop with 
probability $p$, then they can branch with probability $q$, then again hop, etc. In model B 
particles can hop at any time step with probability $p$ or branch with probability $q$, or stay 
idle with probability $1-p-q$. In model B we have of course the restriction $p+q\leq 1$, 
while no such restriction holds in model A. In order to guarantee that annihilation happens 
only between particles in the same generation, we used two data structures: Two lists $L_1$ 
and $L_2$ of integer particle positions and a 1-d array $S$ of characters containing the 
occupancy state of each lattice site. Assume that 
at time step $\tau$ the particles are at sites $i_1,\ldots i_N$ stored in $L_1$, and $S$ is empty. To 
proceed to the next time step, we first go through $L_1$ and update $S$ at the positions $i_1$ 
and $i_n\pm1$, taking into account annihilation. Then we go again 
through $L$, check the values of $S[i_n]$ and $S[i_n\pm1]$. If they are non-zero, the 
corresponding positions are written into $L_2$ and $S$ is cleared at this position. Finally,
$L_1$ is replaced by $L_2$ and we proceed to the next time step. More precisely, this 
algorithm applies to model B. For model A each time step is composed of two half steps.

\begin{figure}
\begin{center}
\includegraphics[width=0.5\textwidth]{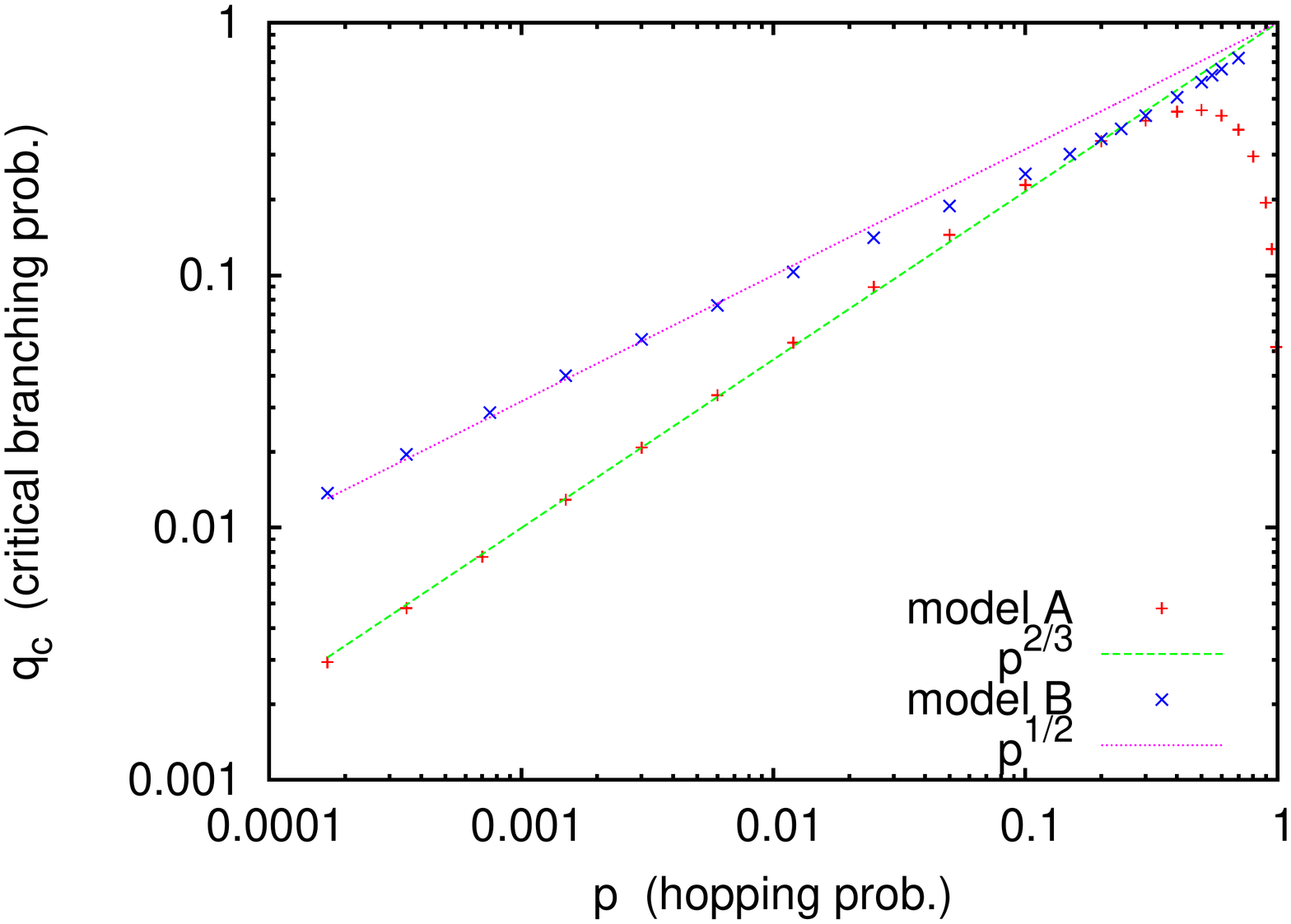}
\caption{(Color online) Log-log plots of critical branching probabilities against hopping 
   probabilities for models A and B. Symbols represent Monte Carlo results whose errors are 
   much smaller than the symbol sizes. Straight lines represent scaling for $p\to 0$.}
   \label{fig_qc}
\end{center}
\end{figure}

We study these models by simulations where we started with two particles at adjacent sites,
and followed their evolution until the population dies or until a prefixed number $T$ of 
time steps is reached. We typically used $10^5 \leq T \leq 5\times 10^7$, with larger $T$ corresponding 
to smaller values of $p$ and $q$. For each value of $p$ we searched for that value of $q$ where 
the process is critical. For this we simply monitored the average number $N(\tau)$
of particles at time step $\tau$. In addition we measured the probability $P(\tau)$ that the process has 
not yet died and the average squared distance $R^2(\tau)$ of the particles from the origin. 
Since it is known that at criticality $N(\tau) \sim \tau^\eta$ with $\eta = 0\pm 0.001$
\cite{Jensen94,zhong95}, it is easy to find the 
critical point with a relative error $<10^{-3}$ by using $\sim 10^5 - 10^7$ runs for each $p$.

Results are shown in Fig.~1. For $q = O(1)$ we found also $p= O(1)$ (except for model A, 
where it seems that $p\to 0$ for $q\to 1$). This is not surprising. It is also not surprising 
that $q\to 0$ when $p\to 0$. But we definitely do not see that the ratio $q/p$ stays finite
when $p\to 0$, as expected in a ``normal" continuous time limit. Rather we find power 
laws
\be
   q \sim p^\alpha
\ee
when $p\to 0$, with $\alpha = 0.50(2)$ for model A and $\alpha = 0.67(2)$ for model B.
We conjecture that the exact exponents are 1/2 and 2/3.

This immediately explains why Sudbury \cite{sudbury90} found only a subcritical phase in the continuum
time limit: To see critical behavior in the limit $p\to 0$ at physical time scales 
corresponding to non-zero diffusion rate, one would have to take the branching rate 
to infinity. 

\begin{figure}
\begin{center}
\includegraphics[width=0.5\textwidth]{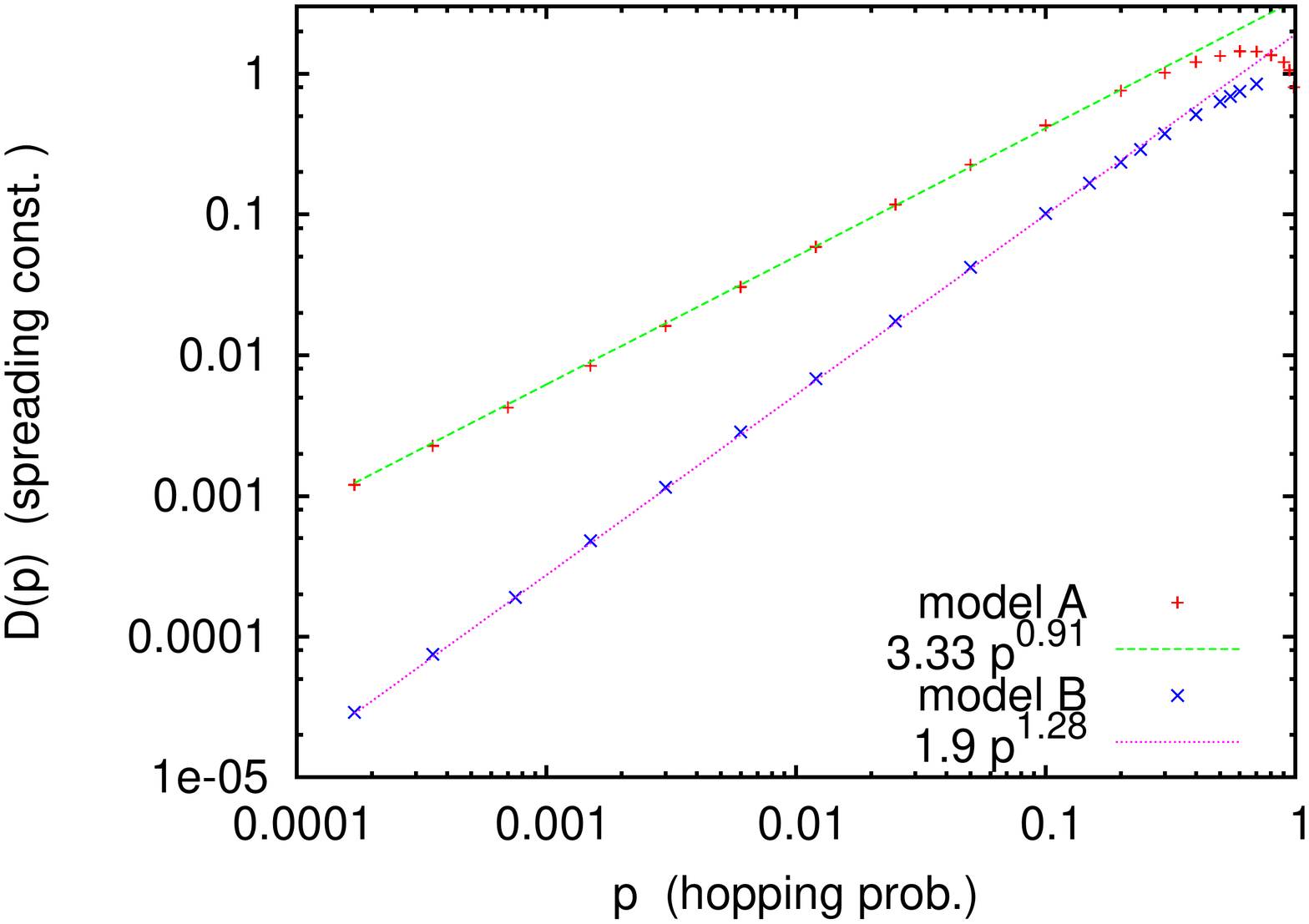}
\caption{(Color online) Log-log plots of spreading constants defined as prefactors in the 
   scaling law $R^2(\tau) \sim \tau^z$ at criticality. Again, errors are
   much smaller than the symbol sizes, and straight lines represent the scaling for $p\to 0$.}
   \label{fig_D}
\end{center}
\end{figure}

For a more precise statement, we obtained from $R^2(t)$ also the speed of 
spreading. In all cases we verified the previous \cite{Jensen94,zhong95} result 
$R^2(\tau) \sim \tau^z$ with $z\approx 1.15$. Therefore we can estimate the $p$-dependent 
`spreading constant' $D(p)$ as 
\be
   D(p) = \lim_{\tau\to\infty} R^2(\tau) / \tau^z.           \label{Dtilde}
\ee
As seen from Fig.~2, $D(p)$ scales for small $p$ as $D(p) \sim p^\sigma$, with 
$\sigma_A = 0.91(3)$ and $\sigma_B = 1.28(3)$.
In a similar way we found that the survival probabilities also scaled with power
laws,
\be
   P(t) \sim p^\rho t^{-\delta}
\ee
where the exponent $\rho$ was the same for both models, $\rho_A = \rho_B = -0.11(1)$.

Eq.~(\ref{Dtilde}) would suggest that we define 
the continuous time spreading constant $D = \lim_{p\to 0}  p^{-\sigma} D(p)$. This
in turn would suggest that we have to define {\it physical} time $t$ as 
$t = p^{\sigma} \tau$, if we want to have $R^2(t) \approx D t^z$ with a finite value of $D$.

Unfortunately, this redefinition of time would lead to singular survival probabilities, 
because $\sigma \neq \rho$ for both models. Therefore, in order to obtain finite renormalized 
parameters we also have to rescale space.

We presented these details in spite of the fact that there exists no fundamental reason 
for using a continuous time limit for the pc-BARW, and although more complicated continuous 
time models in its universality class are known \cite{Jensen94,Menyhard,zhong95}. But there
might exist models where things are even more complicated, and where one has a good reason 
to prefer a continuous time formulation. The present paper might give an indication of what 
is needed in order to deal with such a situation. 

Apart from that, both models A and B are simpler than any previously proposed realization 
of the pc-BARW, and much faster to simulate than the continuum models of 
\cite{Jensen94,Menyhard,zhong95}. Since the simplicity of the numerically obtained critical 
exponents is still enigmatic \cite{Jensen94} -- in particular, there is no explanation why 
$\eta$ is compatible with being exactly equal to zero --, one might want to redo simulation with 
very high precision. Either one of the models proposed in this paper would be a good candidate.
In preliminary runs of model A with $p=1/2$ we indeed found $z=1.149(1)$ and $\delta=0.2873(8)$ 
which exclude the conjectures $z=8/7$ and $\delta=2/7$ of \cite{Jensen94}, but for $\eta$ our 
results are still compatible with the conjecture $\eta=0$.

Finally, we should stress that the pc-BARW as defined in \cite{sudbury90} has an infinite
annihilation rate. Whenever two particles meet on the same site, they annihilate immediately,
so that double occupancies of sites are avoided. This would not be a natural assumption in any
field theoretic treatment \cite{Tauber,Canet,Benitez}. In view of this, it is suggestive 
that the problems encountered in the present note result simply from the fact that one 
rate was assumed to be infinite. Whether this is true or not is not clear.

I am indebted to Hugues Chat\'e for discussions and for critically reading the manuscript.

\bibliography{mm}

\end{document}